\documentclass[
    ,final            
  ]
  {aipproc}

\layoutstyle{6x9}

\begin{document}

\title{Phase Synchronization and invariant measures in sinusoidally perturbed chaotic systems}

\author{M. S. Baptista}{
  address={$^1$ Universit{\"a}t Potsdam, Institut f{\"u}r Physik
Am Neuen Palais 10, D-14469 Potsdam, Deutschland}
}

\author{T. Pereira, J. C. Sartorelli, I. L. Caldas}{
  address={Instituto de F\'\i sica, Universidade de S\~ao Paulo,\\
C.P. 66318, 05315-970 S\~ao Paulo, SP, Brasil}}

\author{$^1$ J. Kurths}{
  address={}
  ,altaddress={} 
}

\begin{abstract}

We  show that, in periodically perturbed chaotic systems,  Phase Synchronization appears,
associated to a special type of stroboscopic map, in which not only 
averages quantities are equal to invariants of the perturbation, 
the angular frequency, but also it exists a very large number of non-transient 
transformations, possibly infinity. In cases where there is not phase
synchronization there is either only transitive transformations on the attractor, or 
a finite number of non-transitive transformations. 
We base our statements in experimental and numerical results from 
the sinusoidally perturbed Chua's circuit. 

\end{abstract}

\maketitle 


\section{Introduction}

In observing a  dynamical system it  is important to measure
physical quantities that are  time invariant. So, different
experiments, considering  different sampling time  observations, still
produce  the same quantity. In chaotic  isolated flows, the most basic
invariant  is the  chaotic   attractor. Representing  $\mathcal{X} \in
\Re^d$ to   be the chaotic   attractor, generated  by  the flow $\xi$,
invariance implies that  $\xi^t(\mathcal{X}) = \mathcal{X}$. For
many  experimental practicle reasons, offently   one observes not  the
attractor,   but measures   of  a subspace   of  it,  for  example the
probability measure  of  a subspace $\mathcal{A}  \in   \mathcal{X}$, defined as
$\rho(\mathcal{A})    =    \int_{\mathcal{A}} d\mu$,   with
$\mu(\mathcal{C})$   being  the   distribution function  of   finding
trajectories in  the  subspace $\mathcal{C}$, and $\rho$ the
probability.   This invariant measure implies $\rho(\mathcal{A})=
\rho[\xi^t(\mathcal{A})]$. A consequence of this  property is that  as
one observes the chaotic attractor for instant times $t$
= $n\tau$, and therefore, for  the discrete set represented by
$\mathcal{D} \in  \Re^d$, $\rho(\mathcal{D})  = \rho(\mathcal{X})$, what
means that the   natural probability density  of an  attractor  can be
calculated  by  discrete  observations  of it,  independently   of the
sampling time. 

This  invariant scenario,  so  important   for experiments,  radically
changes when  one    periodically perturb  a    chaotic system.  Being
$\mathcal{X}^{\prime}$ some subset of $\mathcal{X}$ and $\langle \cdot
\rangle$    representing    some     time    invariant  average     of
$\mathcal{X}^{\prime}$,  and  $\omega$     the angular  frequency  of   the
perturbation, one   can     find two   possible     discrete  dynamics
$\mathcal{D}$ with respect to natural measure.  For sufficiently small
perturbation amplitudes,  it  is   still true  that  $\rho(\mathcal{D})
= \rho(\mathcal{X})$,  however for sufficiently large perturbation
amplitudes   $\rho(\mathcal{D}) \neq \rho(\mathcal{X})$,
which means that the natural  measure cannot be obtained for arbitrary
sampling observations. 

However, in a   perturbed system, there  are  still measures that  are
invariant under different time scales and different initial conditions
on  subspaces of the  attractor. One  example of such a
measure is the ratio between the growing of  the phase with respect to
the time.  As  we will see,  when there is Phase Synchronization  (PS)
between  the chaotic system and the  perturbation, this ratio is equal
to the angular frequency of the perturbation. As a consequence, we can
find a  large  number of  discrete  subsets of  the chaotic attractor,
$\rho(\mathcal{D})$,    such   that    $\rho(\mathcal{D}) \neq
\rho(\mathcal{X}^{\prime})$. When there is   not phase synchronization,   either
$\rho(\mathcal{D}) = \rho(\mathcal{X}^{\prime})$, or  there   is
only a finite number of discrete subsets $\rho(\mathcal{D})$. 

In a sinusoidally perturbed  chaotic system Phase Synchronization $PS$
\cite{epa,rose,reviews}  exists, between  a subspace of $\mathcal{X}$ and
the  perturbation if the following two  conditions Eqs. (\ref{PS}) and
(\ref{NC}), apply: {\small 
\begin{equation}
|\phi(t)-r \omega t|<c, \label{PS} 
\end{equation}}
where the phase $\phi(t) = \phi [ \mathcal{X}(t)]$, and $c\in \Re $ is
a  constant chosen according  to  the particular  system  studied.  We
consider  $r$   to be a  rational   number,  although it  can also  be
irrational  as shown in \cite{murilo}, and $\omega$ is the phase of
the chaotic attractor \cite{epa,rose,reviews}.  A  necessary condition to $PS$
is {\small 
\begin{equation}
\langle \frac{d\phi(t)}{dt} \rangle - r \omega = 0 
\label{NC}
\end{equation}}
\noindent
where $ \langle - \rangle$ is the  average taken over an infinity time
interval. 

In this work we denote $\phi(\tau_{i})$ to phase at instants $\tau_i$.
The quantity  $\langle  \Delta \phi  (\tau_i)
\rangle  = \langle \frac{\phi(\tau_{i+1})-\phi(\tau_i)}{\tau_i} \rangle$ is
an invariant, which means that it has  the same value independently of
the initial  conditions,  and the  time interval  for  its calculation
$\tau_i$. Whenever Eqs. (\ref{PS}) and (\ref{NC}) are satisfied, thus it
is also true that 

\begin{equation}
\langle \Delta \phi(\tau) \rangle =  \omega
\label{ergodic_condition}
\end{equation}
\noindent

{\bf The  sinusoidally perturbed Chua's  circuit:} \ It is represented
by:        
\begin{eqnarray}
C_{1}     \frac{dx_{1}}{dt}=g(x_{2}-x_{1})-i_{NL}(x_{1}) \\
C_{2}\frac{dx_{2}}{dt}=g(x_{1}-x_{2})+x_{3} \\
L\frac{dx_{3}}{dt}=-x_{2}-V\sin {(2\pi ft)}
\end{eqnarray}
Where $x_{1}$, $x_{2}$,
and $x_{3}$ represent, respectively, the tension across two capacitors
and the current   through  the inductor (See  \cite{murilo1}  for more
details), where the $f$ \ is the frequency  and \ $V$ the amplitude of
the perturbation and they  are  the control parameters.  The circuit's
attractors are setup by the $g$ parameter,  and here we adjusted it to
obtain a R\"{o}ssler-like chaotic  attractor.  The perturbation has  a
well defined phase $2\pi ft=\omega t$. 

\begin{figure}
\includegraphics[height=.3\textheight]{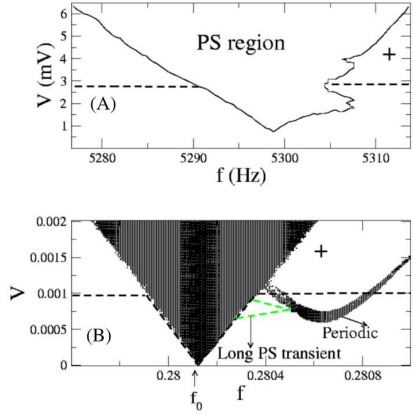}    
\caption{(a)  Experimental  $PS$ parameter   space.  b)  Simulated
$PS$  parameter space,   with  parameters    $g=0.574$,  $C_1$=10,
$C_2$=6, and $L$=6. Black points  represent parameters for  which Eq.
(\ref{PS}) is  satisfied.  In both figures, the  horizontal axis
represents the perturbing frequency  $f$ and the  vertical axis  its
amplitude $V$. Variables  in (b) are dimensionless and  $f_0$ is the
frequency of the non perturbed circuit.}
\label{firenze_fig1} 
\end{figure}

Next   we present some  formalism  using  elements of the  Ergodic  Theory.
Given  a   group  of diffeomorfisms   $F_t$, that   is  the flow,   we
call$\mathcal{X}$  its $\omega$-limit   set.  If   $\mathcal{X}$ is
chaotic then $F_t(\mathcal{X})$ is   mixing,   transitive and ergodic.
However, others transformations applied on $\mathcal{X}$, as
stroboscopic mapping, may not possesses all these properties. 

The notion of stroboscopic mapping  of a flow $F_{t}(\mathcal{X})$ can
be          formalized        defining           a      transformation
$T^{\tau_{i}}:\mathcal{X}\rightarrow \mathcal{X}$, a discretization of
the attractor.  Given a  point $x_{0}\in \mathcal{X}$ we have $T^{\tau
_{i}}(x_{0})=F_{\tau  _{i}}(x_{0})$, so    we  constructed  the  orbit
$\{x_{0},T^{\tau  _{1}}(x_{0}),T^{\tau  _{2}}(x_{0}),\ldots   ,T^{\tau
_{N}}(x_{0}),\ldots  \}$.  We  call $\mathcal{D}$  the $\omega -$limit
set generated by $T^{\tau_i}$. 

A  subspace  $\mathcal{A}$ of   $\mathcal{X}$, is  said   invariant by
transformation        $T^{\tau_{i}}$           if             $T^{\tau
_{i}}(\mathcal{A})=\mathcal{A}$,  $\forall \tau _{i}$,  and a physical
measure $\rho $ is  said  invariant ($T^{\tau _{i}}$-invariant) on   $
\mathcal{A}$ if $\rho (T^{-\tau _{i}}(\mathcal{A}))=\rho (\mathcal{A})$.
$T^{\tau_i}$        is      topologically        transitive         in
$\mathcal{A}$\cite{wiggins} if for  any   two open sets    $B,C\subset
\mathcal{A}$, {\small 
\begin{equation}
\exists \tau _{i}\mbox{\ \  }/\mbox{\ \ }T^{\tau  _{i}}(B) \cap C \neq
\emptyset 
\label{transitive}
\end{equation}
\noindent
} 

{\bf Definition 1}:  Two  sets  $\mathcal{A}$ and $\mathcal{B}$    are
equivalent,  $\mathcal{A}\equiv    \mathcal{B}$,  if $\forall x_{i}\in
\mathcal{A}$, a set  $\mathcal{C}$ can be  constructed by the union of
open sets $B_{\ell }(x_{i})$, open $R^{d}$ volumes centered at $x_{i}$
with  length  $\ell  $,  such  that   $\forall  y_{i}\in  \mathcal{B}$
$\Longrightarrow  $ $y_{i}\in \mathcal{C}$, and $\mathcal{A}\not\equiv
\mathcal{B}$ if $y_{i}\notin \mathcal{C}$. 

In \cite{baptista:2004}  it was shown  that $PS$ is  always associated
with special    non-transitive transformations  that
decomposes  the attractor in discrete chaotic  subsets. To see this we
introduce some notations  and  definitions from  \cite{baptista:2004}.
If  there is  a  non-transitive transformation  $T^{\tau}$, thus,  the
discrete set $\mathcal{D}$ is a basic set  if $\mathcal{D} \not \equiv
\mathcal{X}$, and it is formed by a finite union of open invariant and
{\it minimal} sets  $\mathcal{D}^1,  \ldots , \mathcal{D}^N$,  and for
each   $\mathcal{D}^i$ there is a   $x \in \mathcal{D}^i$ which, under
$T^{\tau_N}(\mathcal{X}^j)$ has a dense orbit on $\mathcal{D}^i$. And,
a  minimal  set $D^i$   cannot   be   further decomposed.   Also,   if
$\mathcal{X}$   can    be       decomposed     into   a     collection
$\mathcal{D}=\mathcal{D}^0,\mathcal{D}^1,\ldots  \mathcal{D}^{P-1}$ of
subsets      of $  \mathcal{X}$,       with  $P\geq   1$,  such   that
$T(\mathcal{D}^i)\subseteq    \mathcal{D}^{i+1        (modP)}$,    and
$T^{P}(\mathcal{D}^{i})\subseteq \mathcal{D}^{i}$,   we  refer to each
minimal set   $\mathcal{D}^i$  as a \textit{recurrent  decomposition}.
The number of sets $P$ is  called \textit{length} of the decomposition
\cite{JB}. 

In Fig. \ref{firenze_fig1}(A), PS of the circuit with the perturbation
is experimentally detected whenever there is a subset  $D$, done by an
observation      time    $\tau_i=i\frac{1}{f}$,     such          that
$\mu(\mathcal{X}^{\prime})$     is very        different          from
$\mu(\mathcal{D})$. In  practice,  we  detect a subset  $\mathcal{D}$,
that contains  points concentrated in an  angular section smaller than
2$\pi$.   Simulation  is shown  in  Fig.  \ref{firenze_fig1}(B), where
black points  represent perturbing parameters  for which condition for
$PS$ is satisfied,  where we assumed that in  Eq.  (\ref{PS}) $r$ = 1,
$c_s$   =   $2\pi$. Also,  both    sides of   the   $PS$  region   are
approximately symmetric. In  this figure, the triangular shaped region
denoted by the light gray dashed line represents the region where long
$PS$ transient happens.   This  transient region  is connected with  a
period-3 region. The dashed line in this figure indicate the parameter
regions for which we  can find $\mathcal{D} \not \equiv  \mathcal{X}$,
with equivalence as defined in Definition 1. 

\begin{figure}
\includegraphics[height=.3\textheight]{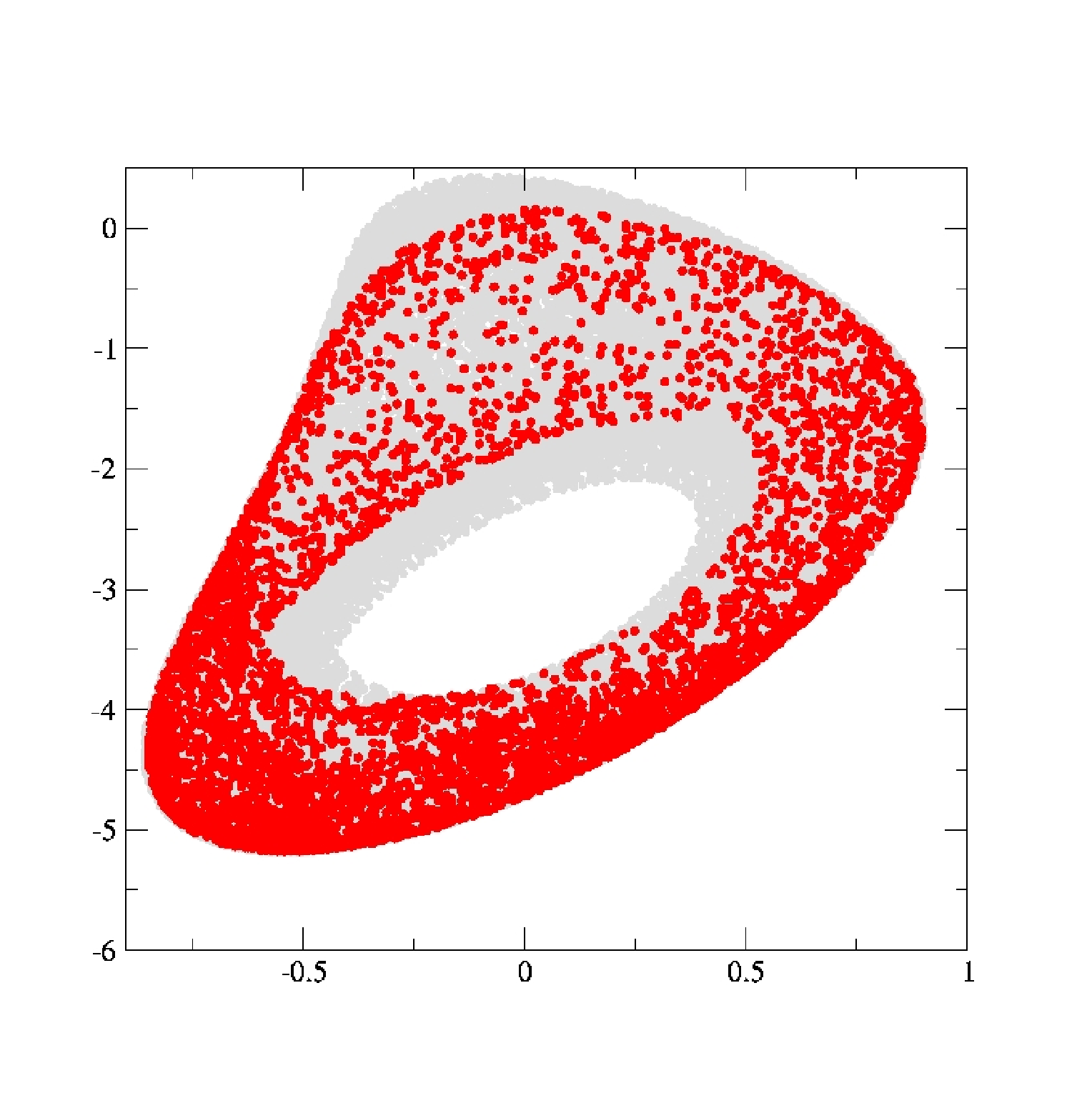}
\includegraphics[height=.3\textheight]{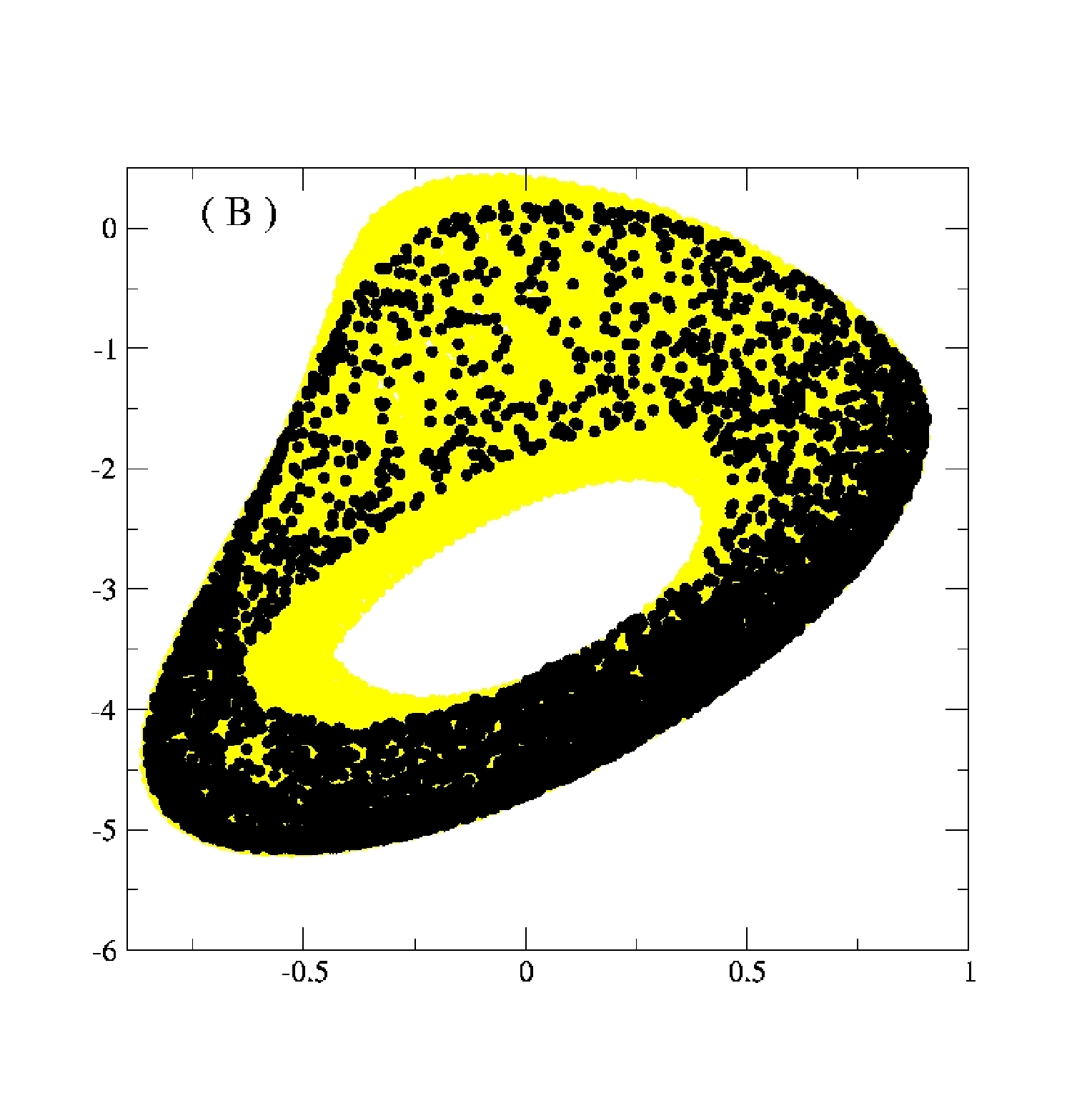}
\end{figure}

\begin{figure}
\includegraphics[height=.3\textheight]{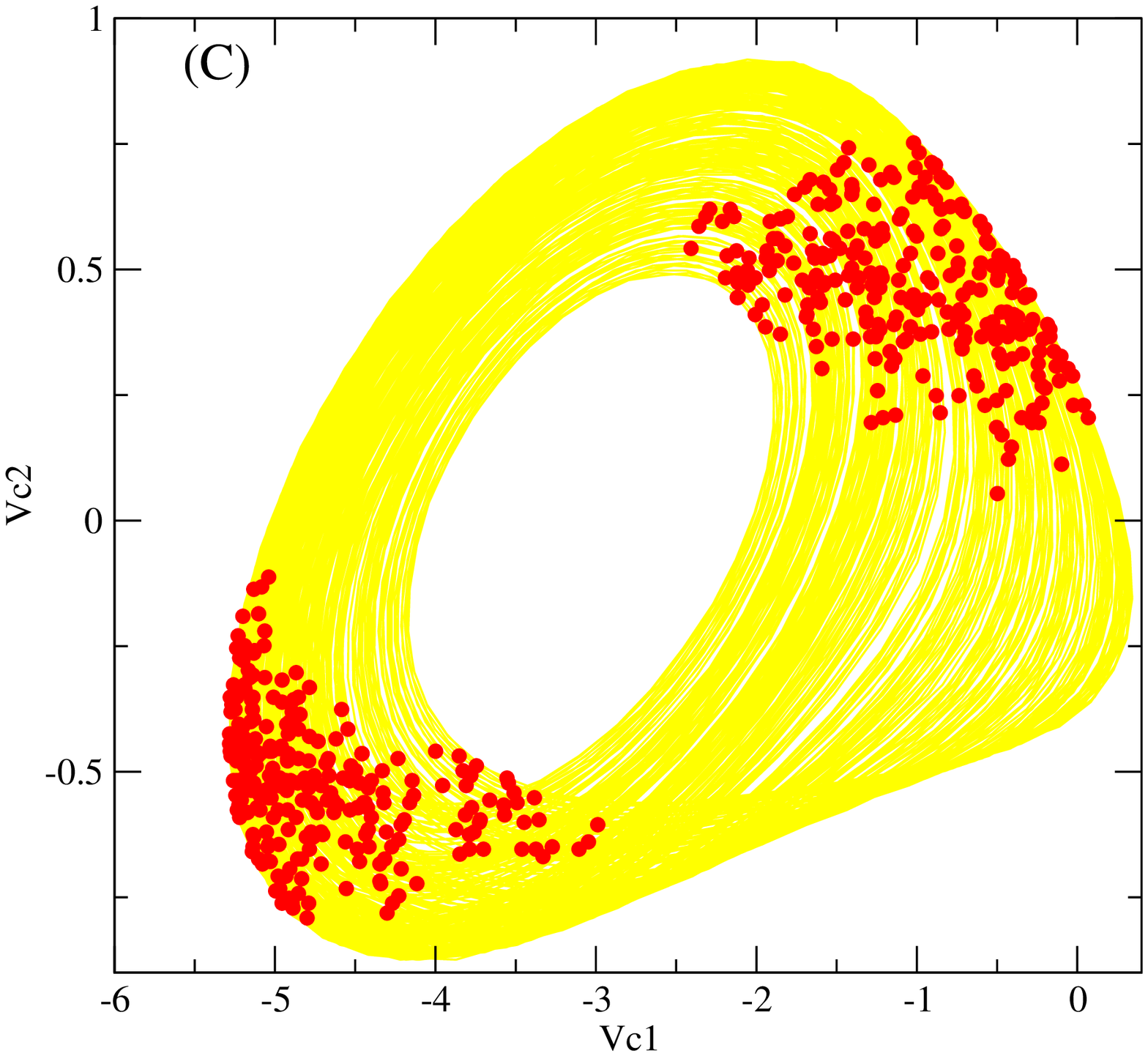}
\includegraphics[height=.3\textheight]{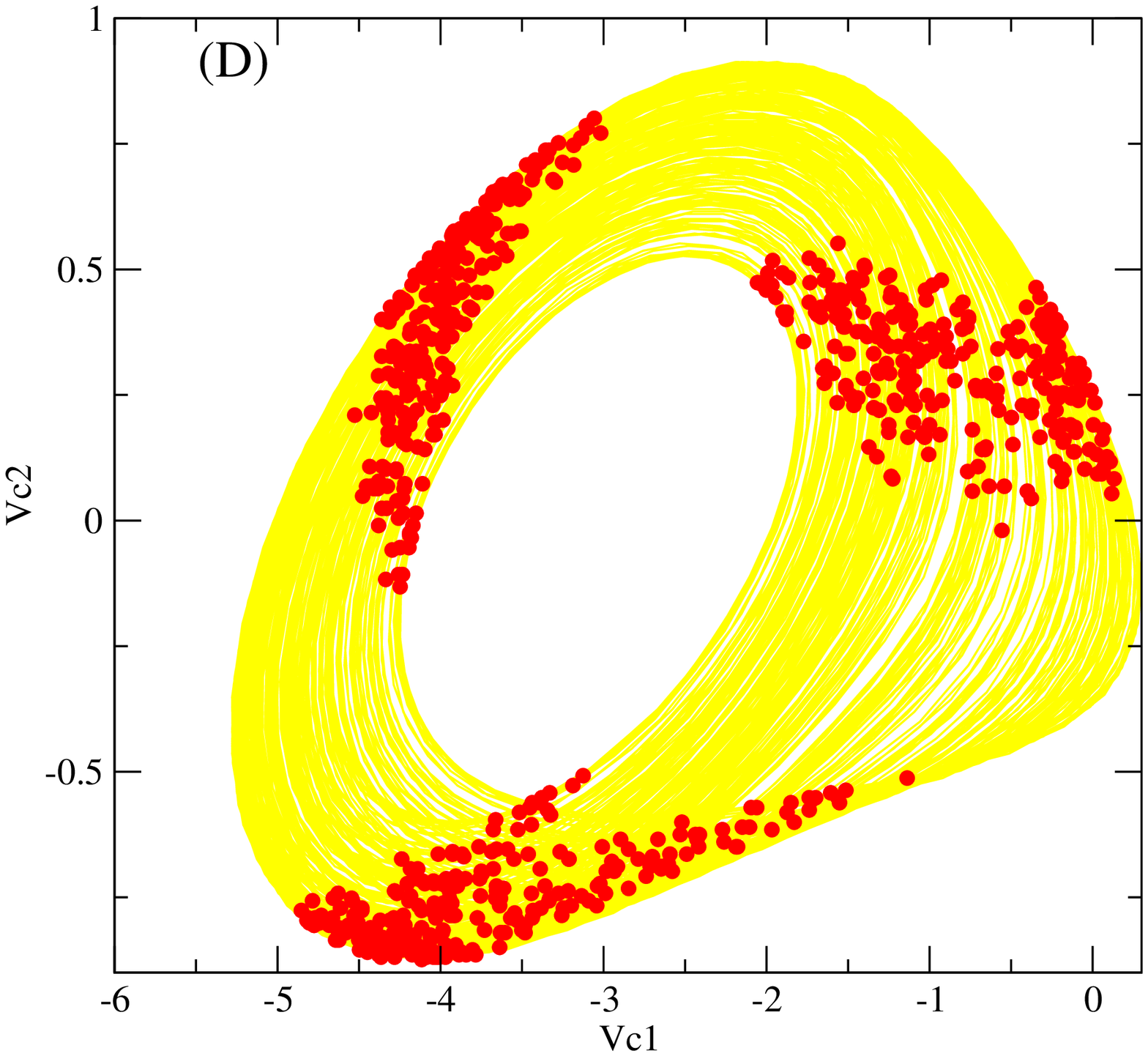}
\caption{Filled balls shows projections  of  the discrete basic  sets,
and  filled  lines, a projection  of the  set $\mathcal{X}$, i.e., the
sets $\mathcal{X}^{\prime}$.  In (A-B), we show  two minimal sets of a
length-2  basic set, constructed for a  time  series given by $i \tau,
(2i+1)/2 \tau,  (i+1)  \tau,  ldots$.    In  (A)  we plot    the   set
$\mathcal{D}_0$, for points observed in the time series $i \tau, (i+1)
\tau, (i+2)  \tau$,  and in  (B)  we plot the set  $\mathcal{D}_1$ for
points observed for the   time  series $(2i+1)/2 \tau,  (2i+3)/2  \tau,
(2i+5)/2  \tau, \ldots$.  In (C), we  show  a length-2 basic set,  for
parameters within  the  phase synchronization region  (V=2 mV and f=5294
Hz),
for  time series constructed  as in Figs. (A)-(B).  In  (D), we show a
length-3 basic  set, constructed for  a time series  $i \tau, (2i+1)/3
\tau, 2(2i+1)/3 \tau, (i+1) \tau, \ldots$.} 
\label{firenze_fig2} 
\end{figure}

In Fig.  \ref{firenze_fig2} we show  in (A) one  minimal set $D_0$ and
(B) another minimal  set  $D_1$ of a   length-2 basic set $D$, for   a
situation for parameters for which there is not phase synchronization,
indicated by the cross in Fig.  \ref{firenze_fig1}(A).  In (B) we show
a length-2  basic   set  for  parameters  for  which  there  is  phase
synchronization.  Note   that while  $\mathcal{D}$   is  approximately
equivalent to  $\mathcal{X}$  in (A-B), $\mathcal{D}$  in (C)  as well
$\mathcal{D}$ in (D) is absolutely not equivalent to $\mathcal{X}$. 

In  conclusion,   we  show that   phase  synchronization  implies  the
existence of many invariant measure associted to sampling of perturbed
chaotic attractor. In PS states  these invariant measure are associted
to    basic sets.   Experimental  and Numerical    examples of   these
statementes are presented of Chua's Circuit. 



\begin{theacknowledgments}
Research partially financed by the Alexander von Humbold fondation,
and by Brasilian agencies FAPESP and CNPq. 
\end{theacknowledgments}


\bibliographystyle{aipproc}   

\bibliography{sample}

\begin{thebibliography}{99}

\bibitem{baptista:2004} M.S. Baptista, T. Pereira, J.C. Sartorelli and
I.L. Caldas, submitted for publication 


\bibitem{murilo} M. S. Baptista, S. Boccaletti, K. Josi{\'c}, and 
I. Leyva, Irrational Phase Synchronization, to appears in Phys. Rev. E.

\bibitem{murilo1} M. S. Baptista, T. P. Silva, J. C. Sartorelli, and
I. L. Caldas, E. R., Jr., Phys. Rev. E {\bf 67}, 056212 (2003).

\bibitem{epa} E. Rosa, Jr., E. Ott, and M. H. Hess, 
Phys. Rev. Lett. {\bf 80} (1998).

\bibitem{rose}
M.G. Rosenblum, A.S. Pikovsky and J. Kurths, Phys. Rev. Lett. {\bf
76}, 1804 (1996); M. G. Rosenblum, A. S. Pikovsky, and J. Kurths, 
Phys. Rev. Lett. {\bf 78}, 4193 (1997). 

\bibitem{reviews}
A. Pikovsky, M. Rosenblum and J. Kurths, {\it Synchronization: A
Universal Concept in Nonlinear Sciences}, (Cambridge University
Press, 2001); S. Boccaletti, J. Kurths, G. Osipov, D. Valladares
and C. Zhou, Phys. Rep.  {\bf 366}, 1, (2002).

\bibitem{JB} J. Banks, J. Ergod. Th. and Dynam. Sys. {\bf 17} 505
(1997).

\bibitem{wiggins} S. Wiggins, {\it Introduction to Applied Nonlinear 
Dynamical Systems and Chaos} (Springer, New York 1996).










\end{thebibliography}

\IfFileExists{\jobname.bbl}{}
 {\typeout{}
  \typeout{******************************************}
  \typeout{** Please run "bibtex \jobname" to optain}
  \typeout{** the bibliography and then re-run LaTeX}
  \typeout{** twice to fix the references!}
  \typeout{******************************************}
  \typeout{}
 }

\end{document}